\documentclass[11pt]{article}
\usepackage{moriond,epsfig}
\usepackage{amssymb}
\def\Journal#1#2#3#4{{#1} {\bf #2}, #3 (#4)}

\def\NIM{\em Nucl. Instrum. Methods}

\def\PLB{{\em Phys. Lett.}  B}
\def\PRL{\em Phys. Rev. Lett.}

\def\JPhys{{\em J. Phys.} G}
\def\EPJC{{\em Eur. Phys. J.} C}

\def\be{\begin{equation}}
\def\ee{\end{equation}}
\def\bea{\begin{eqnarray}}
\def\eea{\end{eqnarray}}

\begin{document}
\vspace*{4cm}
\title{Search for the Standard Model Scalar Boson with the ATLAS detector}

\author{ Sandra Kortner on behalf of the ATLAS Collaboration}

\address{Max-Planck-Institut f\"ur Physik, F\"ohringer Ring 6, D-80805 Munich, Germany}

\maketitle\abstracts{
The experimental results of the search for the Standard Model Higgs boson with the ATLAS detector at the Large Hadron Collider are reported, based on a dataset of $pp$ collision data with an integrated luminosity of up to 4.9 fb$^{-1}$ at $\sqrt{s}$ = 7 TeV. The search combines several Higgs boson decay channels in a wide range of Higgs boson masses from 110~GeV to 600~GeV. A Standard Model Higgs boson is excluded at the 95\% confidence level in the mass ranges from 110.0~GeV to 117.5~GeV, 118.5~GeV to 122.5~GeV, and 129~GeV to 539~GeV, while the range from 120~GeV to 555~GeV is expected to be excluded in the absence of a signal. The most significant excess of events is observed around 126~GeV with a local significance of 2.5$\sigma$. The global probability for such an excess to occur in the full searched mass range is approximately 30\%.}

\section{Introduction}

One of the main missing pieces of puzzle in the Standard Model (SM) of particle physics relates to the electroweak symmetry breaking mechanism~\cite{EnglertBrout,Higgs1,Higgs2,GHK} which predicts the existence of a new, yet undiscovered scalar boson, the Higgs boson. Electroweak precision measurements at the LEP, SLD and Tevatron experiments~\cite{lepindirect} set an indirect upper limit on the Higgs boson mass of $m_H<$~152~GeV at the 95\% confidence level (CL). The existence of a Higgs boson in the mass regions with $m_H<$~114.4~GeV and 156~GeV~$<m_H<$~177~GeV is excluded at the 95\% CL by the LEP~\cite{LEP} and Tevatron~\cite{Tevatron} experiments, respectively. In the following, the latest results from direct searches for the SM Higgs boson with the ATLAS detector at the Large Hadron Collider (LHC) are presented. The results superseed the ones published in Ref.\cite{previous_comb} and are based on the full dataset of $pp$ collision data recorded in 2011 with an integrated luminosity of up to 4.9~fb$^{-1}$ at a center-of-mass energy of $\sqrt{s}$=7~TeV.

\section{Search overview}
The search for the SM Higgs boson is performed in the mass range from 110~GeV to 600~GeV combining several search channels summarized in Table~\ref{channel_summary}. The Higgs boson decays into vector boson pairs allow for the search in the entire mass range, while the $H\to~ \gamma\gamma$, $H\to~\tau^+\tau^-$ and $H\to b\bar{b}$ decay modes provide an additional sensitivity for a low-mass Higgs boson ($m_H<$~150~GeV). As opposed to other channels, the $H\to \gamma\gamma$ and $H\to ZZ\to 4\ell$ decay modes are distinguished through their high signal mass resolution. Each search channel is divided into several exclusive sub-channels with different background composition or signal-to-background ratios. 
\begin{table}[!hbtp]
\caption{Individual channels in search for the SM Higgs boson along
with the corresponding number of sub-channels, explored range of Higgs boson masses, integrated luminosity and references to public documentation.\label{tab:exp}}
\label{channel_summary}
\vspace{0.4cm}
\begin{center}
\begin{tabular}{|l|c|c|c|c|}
\hline
Search channel& Nr. of subchannels&  $m_H$ range (GeV)& L (fb$^{-1}$)& Reference\\

\hline
$H\to \gamma\gamma$& 9& 110-150& 4.9& \cite{gg}\\
$H\to \tau\tau$& 12& 110-150& 4.7& \cite{tautau}\\
$VH, H\to bb$& 11& 110-130& 4.7& \cite{bb}\\
$H\to WW^{(*)}\to \ell\nu\ell\nu$& 9& 110-600& 4.7& \cite{lvlv}\\
$H\to ZZ^{(*)}\to 4\ell$& 4& 110-600& 4.8& \cite{4l}\\
$H\to ZZ\to\ell\ell\nu\nu$& 4& 200-600& 4.7& \cite{llvv}\\
$H\to ZZ\to\ell\ell jj$& 2& 200-600& 4.7& \cite{llqq}\\
$H\to WW\to\ell\nu jj$& 4& 300-600& 4.7& \cite{lvqq}\\

\hline
\end{tabular}
\end{center}
\end{table}
The exclusion limits on the Higgs boson production cross section are set based on the mass spectra of the final decay products, using  the modified frequentist approach ($CL_S$)~\cite{cls}. The $p_0$-value, i.e. the probability that the expected background fluctuates as high as the observed number of events or higher, has been evaluated for each hypothesized mass $m_H$ using a frequentist approach~\cite{p0}. The Higgs boson production cross sections and decay branching ratios, as well as the corresponding theory uncertainties are taken into account and are summarized in Ref.~\cite{YellowReport1}. The QCD
scale uncertainties depend on the mass $m_H$ and typically amount to~$^{+12}_{-8}$\% for the dominant gluon fusion ($ggH$) production mode, $\pm$1\% for the production via vector boson fusion (VBF) and in association with vector bosons (VH), and~$^{+3}_{-9}$\% for the associated production with top-quark pairs ($t\bar{t}H$). The uncertainties related to the parton distribution functions (PDF) amount to $\pm$8\% for the predominantly gluon-initiated $ggH$ and $t\bar{t}H$  processes, and $\pm$4\% for the predominantly quark-initiated VBF and VH processes. 


\section{Searches in the low-$m_H$ region (110~GeV - 150~GeV)}

\subsection{$H\to \gamma\gamma$}

\begin{figure}[!htbp]
  \begin{center}
    \psfig{figure=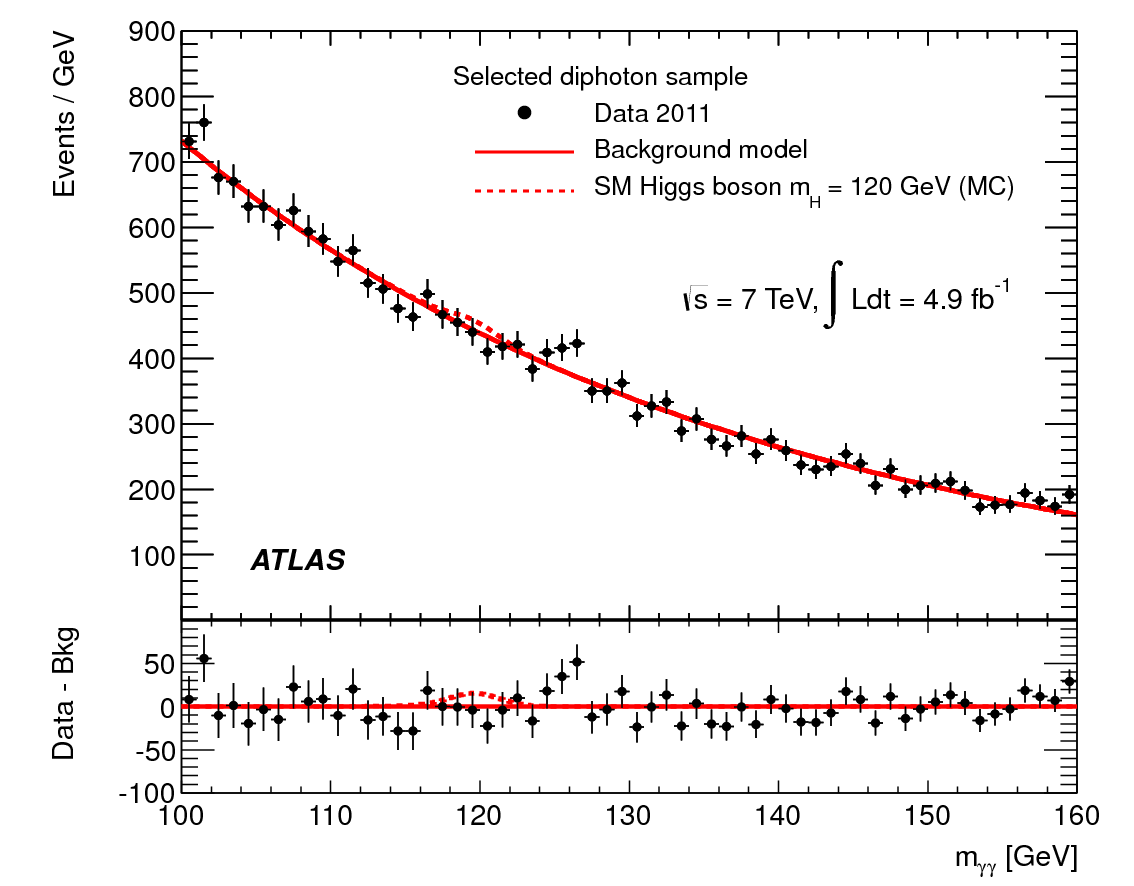,height=60mm}\psfig{figure=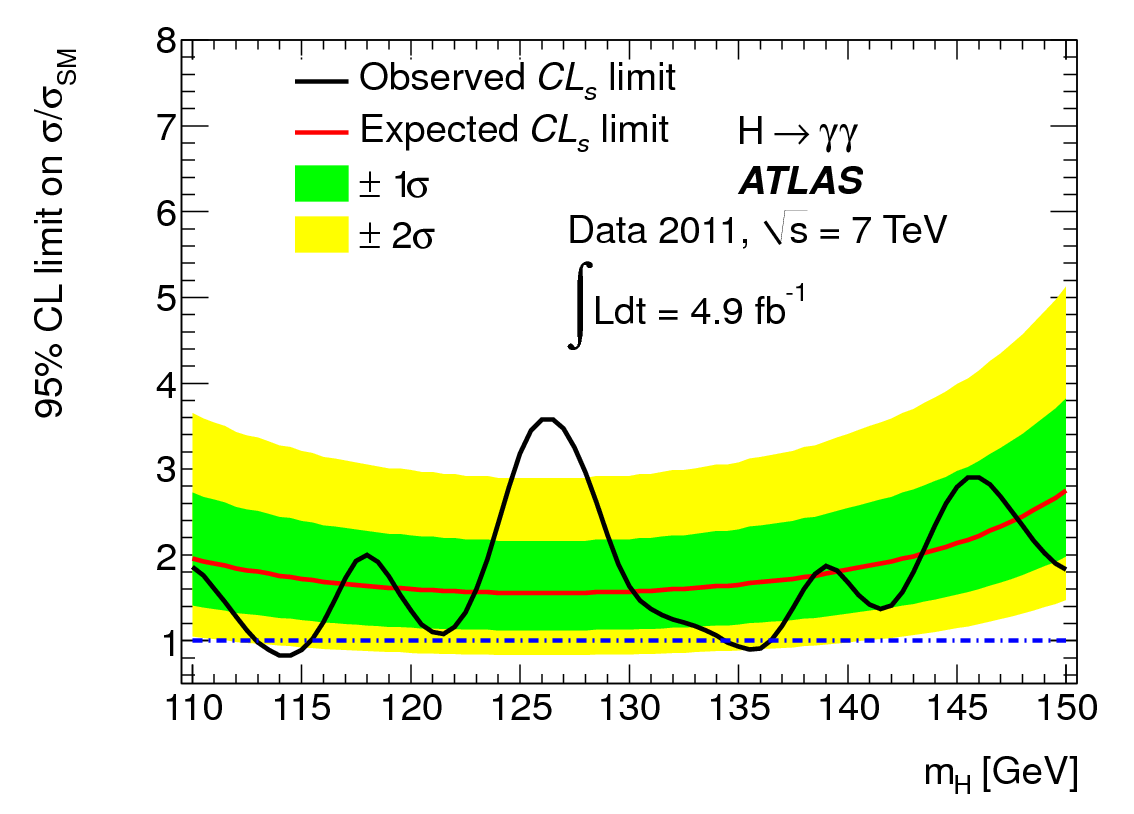,height=60mm}
  \end{center}
  \caption{Left: Invariant diphoton mass distribution, overlaid with the total background from the fit to all sub-categories. The SM Higgs boson expectation for a mass hypothesis of 120~GeV is also shown. Right: Observed and expected 95\% CL limits on the SM Higgs boson production in the $H\to \gamma\gamma$ decay channel normalized to the predicted SM cross section as a function of $m_H$.\label{fig:gg}}
\end{figure}

Despite of a very small branching ratio of about 0.2\%, the $H\to \gamma\gamma$ decay channel provides the highest sensitivity for the Higgs boson search in the mass region $m_H\lesssim$~120~GeV. Selection of two isolated photons with high transverse energy has been optimized for the supression of the reducible photon-jet and jet-jet backgrounds with one or two misidentified jets. The irreducible background with true diphoton events comprises about 70\% of all selected events in the explored mass range from 100~GeV to 150~GeV. The analysis is divided into 9 sub-categories based on the photon pseudorapidity, conversion status and $p_{Tt}^{\gamma\gamma}$, i.e. the component of the diphoton $p_T$ orthogonal to the di-photon thrust axis. Excellent diphoton invariant mass resolution of 1\% to 2\% allows for the search of a narrow di-photon invariant mass peak from the Higgs boson decays on the continuum background. The inclusive invariant diphoton mass spectrum is shown in Figure~\ref{fig:gg} (left) together with the total expected background contribution. The background contributions are estimated separately in each sub-category from the fit of an exponential function to the diphoton invariant mass spectrum, while the signal shape is modelled by the sum of a Crystal Ball and a Gaussian function. 
%
%
The observed and expected exclusion limits at the 95\% CL on the Higgs boson production in units of the SM cross section are shown in Figure~\ref{fig:gg} (right). A SM Higgs boson is excluded at the 95\% CL in the mass ranges of 113~GeV-115~GeV and 134.5~GeV-136~GeV. The largest excess with respect to the background-only hypothesis is observed
at 126.5 GeV with a local significance of 2.8$\sigma$. The corresponding global significance is 1.5$\sigma$ after accounting for the look-elsewhere-effect ($LEE$)~\cite{LEE} in the entire mass range explored with this decay channel.

\subsection{$H\to\tau^+\tau^-$}

\begin{figure}[!htbp]
  \begin{center}
    \psfig{figure=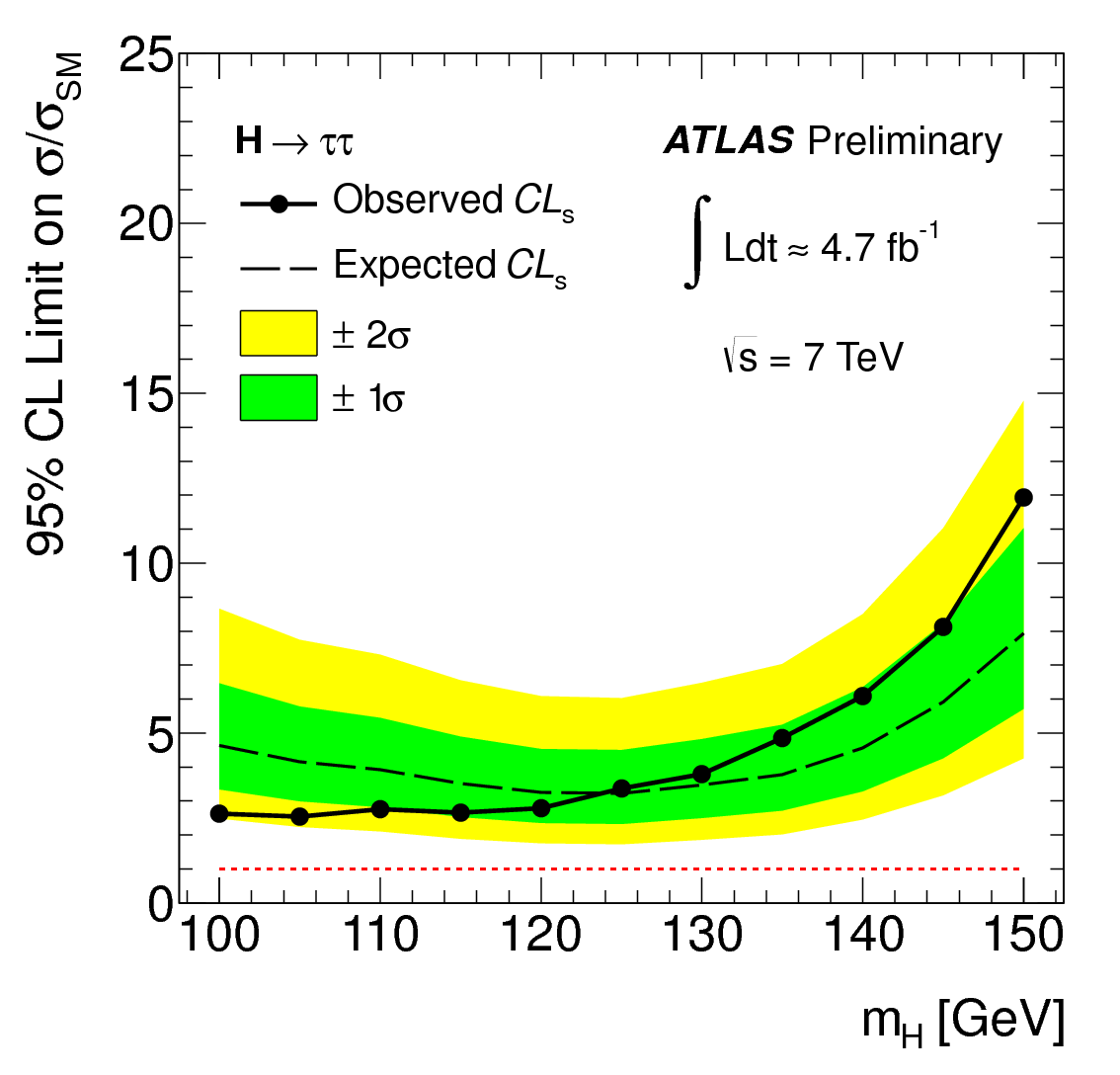,height=60mm}\psfig{figure=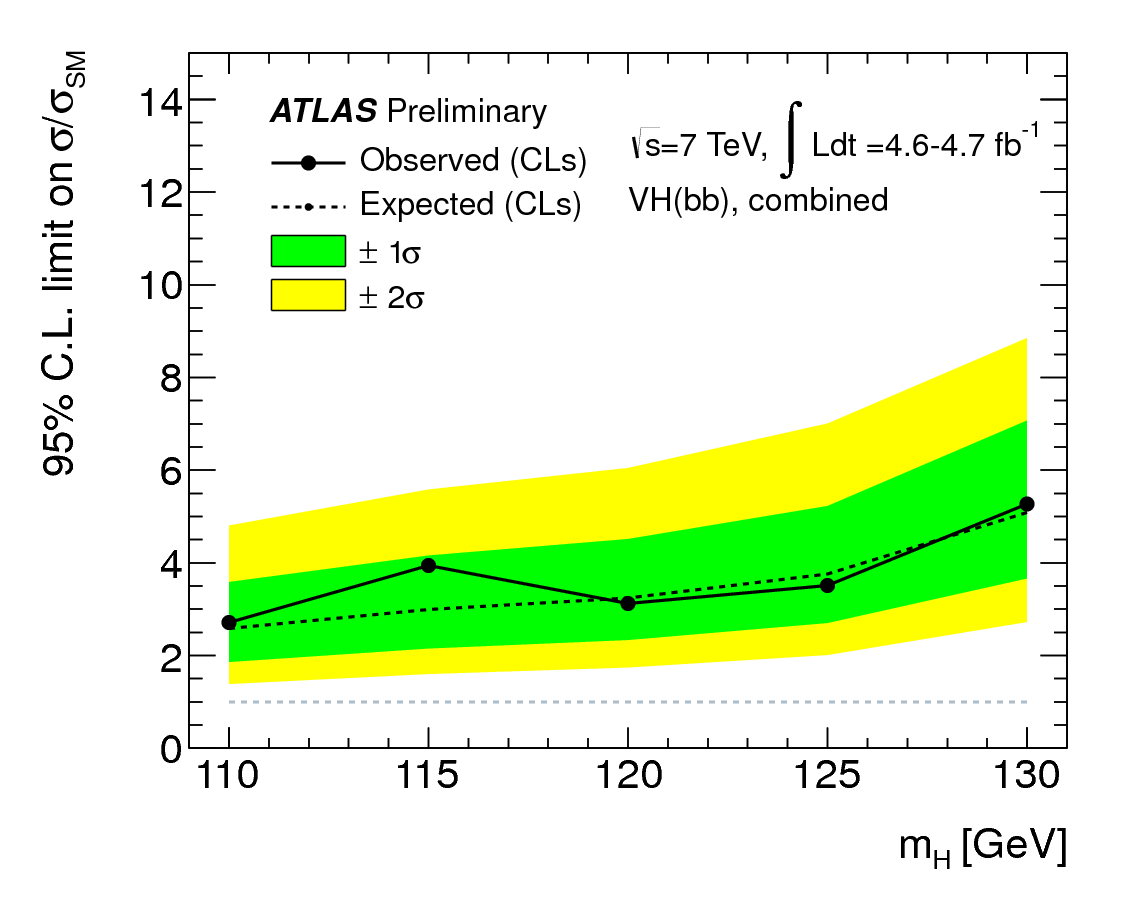,height=60mm}
  \end{center}
  \caption{Observed and expected 95\% CL limits on the SM Higgs boson production in the $H\to \tau^+\tau^-$ (left) and $H\to b\bar{b}$ (right) decay channels normalized to the predicted SM cross section as a function of $m_H$.\label{fig:tt_bb}}
\end{figure}

The mass range from 100~GeV to 150~GeV can also be probed by the less sensitive  $H~\to~\tau^+\tau^-$ decay channel with leptonic and hadronic decay modes of the two tau-leptons resulting in the fully-leptonic $\ell\ell 4\nu$, semi-leptonic $\ell\tau_{had} 3\nu$ and fully-hadronic $\tau_{had}\tau_{had} 2\nu$ final states. For an optimal analysis sensitivity, events are separated into 12 mutually exclusive sub-channels based on the lepton flavour and jet multiplicity, as well on the event topologies characteristic for the VBF and VH signal production mechanisms. The reconstructed mass shape of the dominant $Z~\to~\tau^+\tau^-$ background contribution is estimated by means of an embedding technique in which muons from $Z$ decay events are substituted by simulated
tau-lepton decays. The normalization of this background is obtained from the simulation. Background processes with jets misidentified as leptons or as tau-jets are estimated from the signal-free control data with reversed lepton isolation and tau-jet identification criteria, or samples with a same-sign charge of the two tau-lepton decay products. All other background contributions are estimated from simulation. Depending on the sub-channel the $\tau\tau$ invariant mass distribution used for the limit setting is reconstructed using the effective mass, collinear approximation or a Missing Mass Calculator~\cite{MMC} technique. The 95\%~CL exclusion limits on the SM Higgs boson production are shown in Figure~\ref{fig:tt_bb}. The observed limits vary from 2.5 to 11.9 times the predicted SM cross section over the entire explored mass range. 

\subsection{$VH, H\to b\bar{b}$}

In order to supress the extremely large multijet background contribution, the VH production mode is used to explore the $H\to b\bar{b}$ decay channel in the mass range from 100~GeV to 130~GeV. The search is divided into 11 sub-channels based on the decay mode of the associated vector boson and its transverse momentum: $\ell\nu b\bar{b}$ with four $p_T^W$ bins, $\ell\ell b\bar{b}$ with four $p_T^Z$ bins and $\nu\bar{\nu} b\bar{b}$ with three bins of the mising transverse energy. Since the Higgs and the vector boson tend to recoil away from each other, the highest sensitivity is reached in bins with highest transverse momenta of the vector bosons.  The dominant backround contributions originate from the $Z$+jets, $W$+jets, $t\bar{t}$, diboson and multijet processes. The shape of all but the last background process is obtained from the simulation, while the dedicated control data samples are used to estimate the shape of the multijet contribution and the normalization of all mentioned processes. The exclusion limits shown in Figure~\ref{fig:tt_bb} are set based on the di-$b$-jet invariant mass distribution. The observed upper limits at the 95\% CL on Higgs boson production  vary from of 2.7 to 5.3 times the SM cross section in the entire mass range explored by this channel.

\section{Searches in the high-$m_H$ region (200~GeV - 600~GeV)}

A high-mass Higgs boson will predominantly decay into a pair of vector bosons. 
Final states with a non-leptonic decay of one of the two vector boson have an advantage of a higher decay rate with respect to fully leptonic channels. At the same time, the  backround level can still be controlled using the invariant mass constraint of both vector bosons.
 
\begin{figure}[!htbp]
  \begin{center}
    \psfig{figure=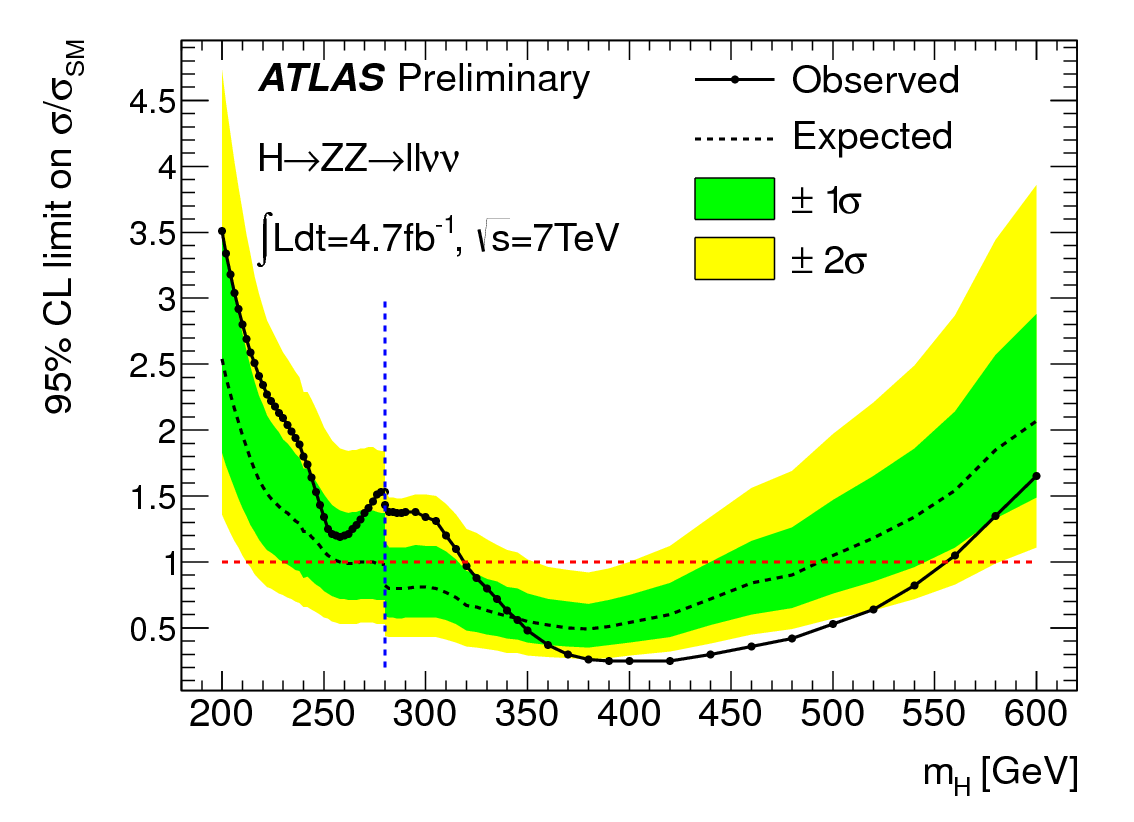,height=60mm}\psfig{figure=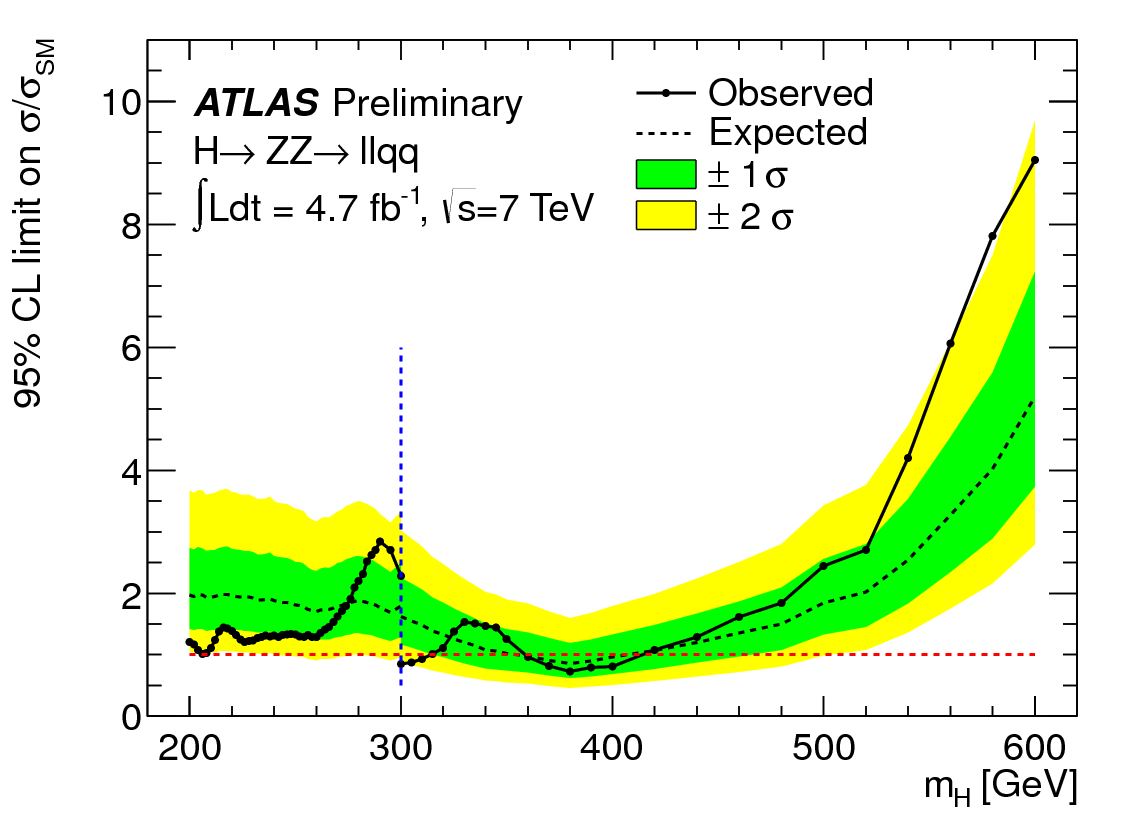,height=60mm}
  \end{center}
  \caption{Observed and expected 95\% CL limits on the SM Higgs boson production in the $H\to ZZ\to \ell\ell\nu\nu$ (left) and $H\to ZZ\to \ell\ell jj$ (right) decay channels normalized to the predicted SM cross section as a function of $m_H$.\label{fig:llvv_llqq}}
\end{figure}

The $H\to ZZ\to \ell\ell\nu\nu$ decay channel provides the highest sensitivity in the high-$m_H$ region due to the high transverse momenta of leptons and neutrinos from the $Z$ boson decays. The search is performed separately in  $ee\nu\nu$ and $\mu\mu\nu\nu$ sub-channels, both separated for low and high pile-up environment to account for the different resolution of the missing transverse energy. The invariant mass of the pair of the charged leptons is required to agree with the $Z$ boson mass within 15 GeV. Different selection cuts are applied for searches below and above $m_H$=280~GeV, accounting for the different level of boost of the $Z$ boson from the Higgs boson decay. The main background processes are suppressed mostly by the cuts on the opening angle of the two leptons from the $Z$ boson decay and on the missing transverse energy.
The contribution of the dominant $ZZ$ background is estimated from simulation, while dedicated control data are used for other processes ($WZ$, $t\bar{t}$ and $W/Z$+jets). The transverse mass distribution is used for the limit setting. The observed (expected) exclusion limit at the 95\% CL covers the mass range from 320~GeV to 560~GeV (260~GeV to 490~GeV), as shown in Figure~\ref{fig:llvv_llqq} (left). 

Somewhat lower sensitivity is reached in the $H\to ZZ\to \ell\ell jj$ channel
which is divided into two sub-channels, one with maximum one jet tagged as a $b$-jet and one with two $b$-jets. Candidate events must have a lepton pair and a dijet pair with invariant masses compatible with the $Z$ boson mass. The dominant $t\bar{t}$ and $Z+jets$ backgrounds are suppressed by the cuts on the missing transverse energy and the opening angle between the two jets. The exclusion limit based on the $m_{\ell\ell jj}$ invariant mass distribution is shown in Figure~\ref{fig:llvv_llqq} (right). Expected exclusion limit at the 95\% CL covers the mass range from 360~GeV to 400~GeV, while the observed upper limit excludes the mass range from 300~GeV to 310~GeV and 360~GeV to 400~GeV.

The $H\to WW\to \ell\nu jj$ channel is separated into six sub-channels based on the lepton flavour and the jet multiplicity (0, 1 and 2 jets with a VBF-like topology). The invariant mass of the dijet pair must be compatible with the $W$ boson mass and a mass constraint $m_{\ell\nu}$=$m_W$ is applied to determine the $z$-component of the neutrino momentum. The limit setting is based on the resulting $m_{\ell\nu jj}$ invariant mass distribution. At present the channel is not sensitive enough to exclude the SM Higgs boson in the explored mass range from 300~GeV to 600~GeV at the 95\% CL.

\section{Searches in the full $m_H$ region (110~GeV - 600~GeV)}

The fully leptonic Higgs boson decays into two vector bosons provide 
a good handle for the background processes even in the case that one of the vector bosons is off-shell. These final states can therefore be used for the search in the entire mass range from 110~GeV to 600~GeV.

\subsection{$H\to WW^{(*)}\to \ell\nu\ell\nu$}

\begin{figure}[!htbp]
  \begin{center}
    \psfig{figure=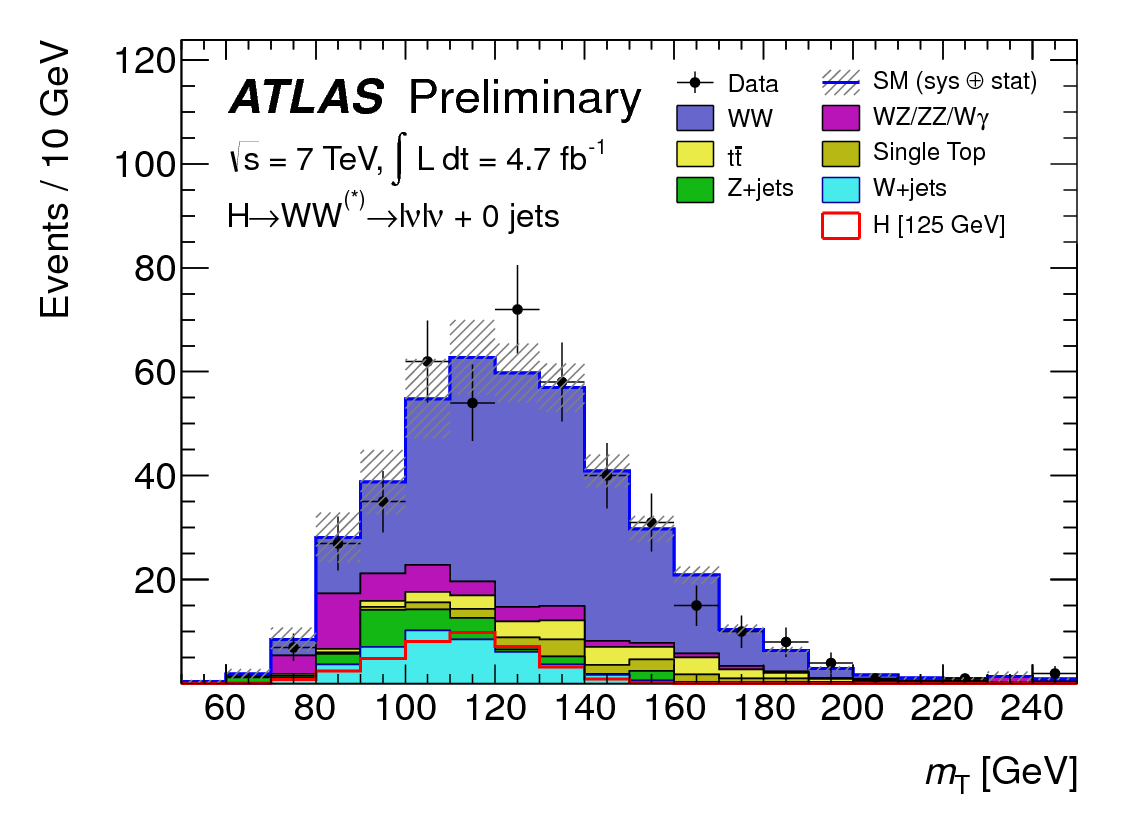,height=60mm}\psfig{figure=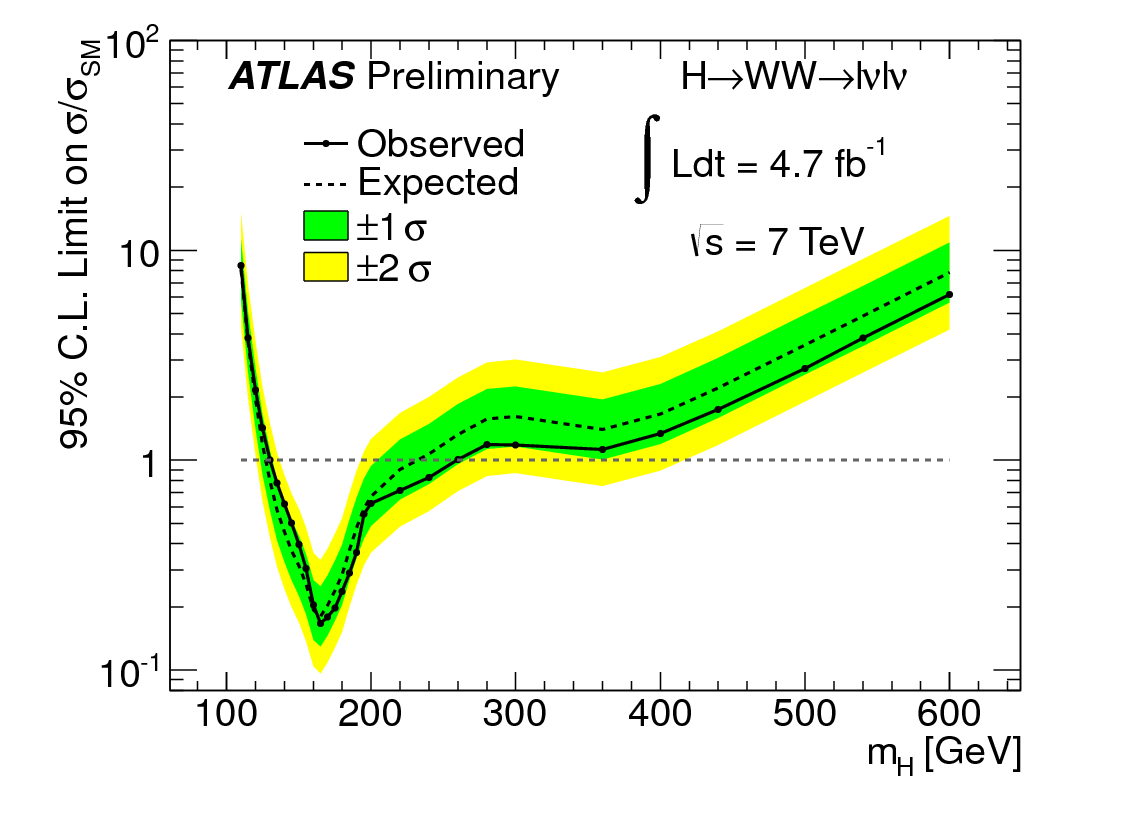,height=60mm}
  \end{center}
  \caption{Left: Transverse mass distribution in the 0-jet $H\to WW^{(*)}\to\ell\nu\ell\nu$ sub-channel, compared to the background expectation.  Right: Observed and expected 95\% CL limits on the SM Higgs boson production in the same channel normalized to the predicted SM cross section as a function of $m_H$. \label{fig:ww}}
\end{figure}

The Higgs boson search in the $H\to WW^{(*)}\to \ell\nu\ell\nu$ channel provides the highest sensitivity in a wide range of hypothesized Higgs boson masses. The analysis is divided in 9 sub-channels: $ee\nu\nu$, $\mu\mu\nu\nu$ and $e\mu\nu\nu$, each separated by the jet multiplicity into final states with no jets, 1 jet or two jets with a VBF-like topology. The channel is characterized by a broad transverse mass distribution $m_T$ for the Higgs signal due to the presence of two neutrinos in the final state. The main reducible backgrounds ($W/Z$+jets, multijets, $t\bar{t}$) are suppressed by the lepton isolation and $b$-jet veto requirement, as well as the cuts on the missing transverse energy. The dominant $WW$ background contribution can be discriminated by means of topological cuts on the invariant dilepton mass and the opening angle between the two leptons. All background contributions remaining after the full selection are normalized using dedicated control regions and extrapolated to the signal region relying on predictions from the simulation. The transverse mass  distribution is used in each sub-channel for the limit setting. As an example,
the $m_T$ distribution in the 0-jet sub-channel is shown in Figure~\ref{fig:ww} (left). The 95\% CL exclusion limit with all sub-channels combined is shown in Figure~\ref{fig:ww} (right). No significant excess of events is observed. A SM Higgs boson
with a mass in the range between 130~GeV and 260~GeV is excluded at the 95\% CL, while the expected exclusion interval ranges from  127~GeV to 234~GeV. 

\subsection{$H\to ZZ^{(*)}\to 4\ell$}

\begin{figure}[!htbp]
  \begin{center}
    \psfig{figure=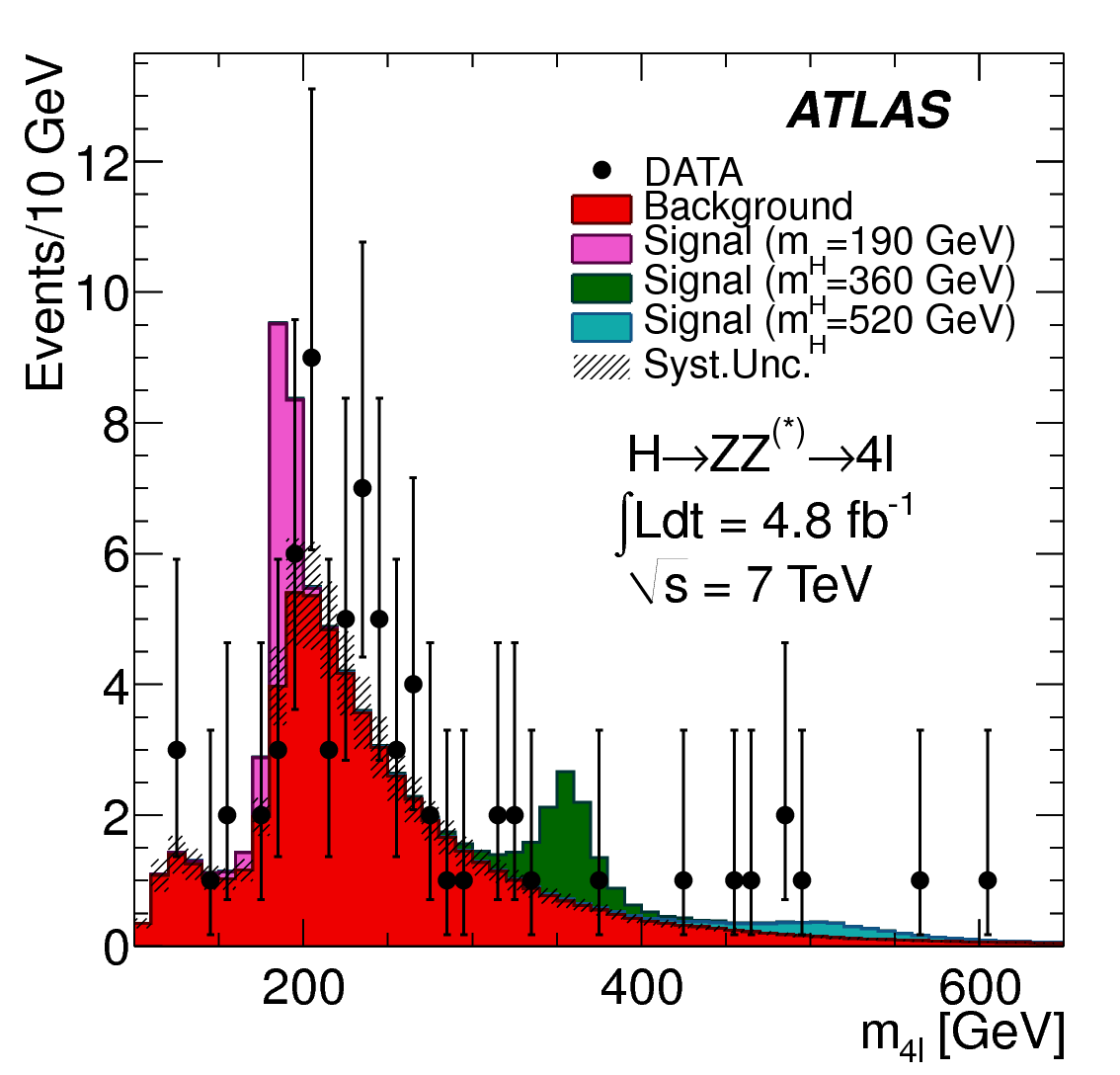,height=65mm}\psfig{figure=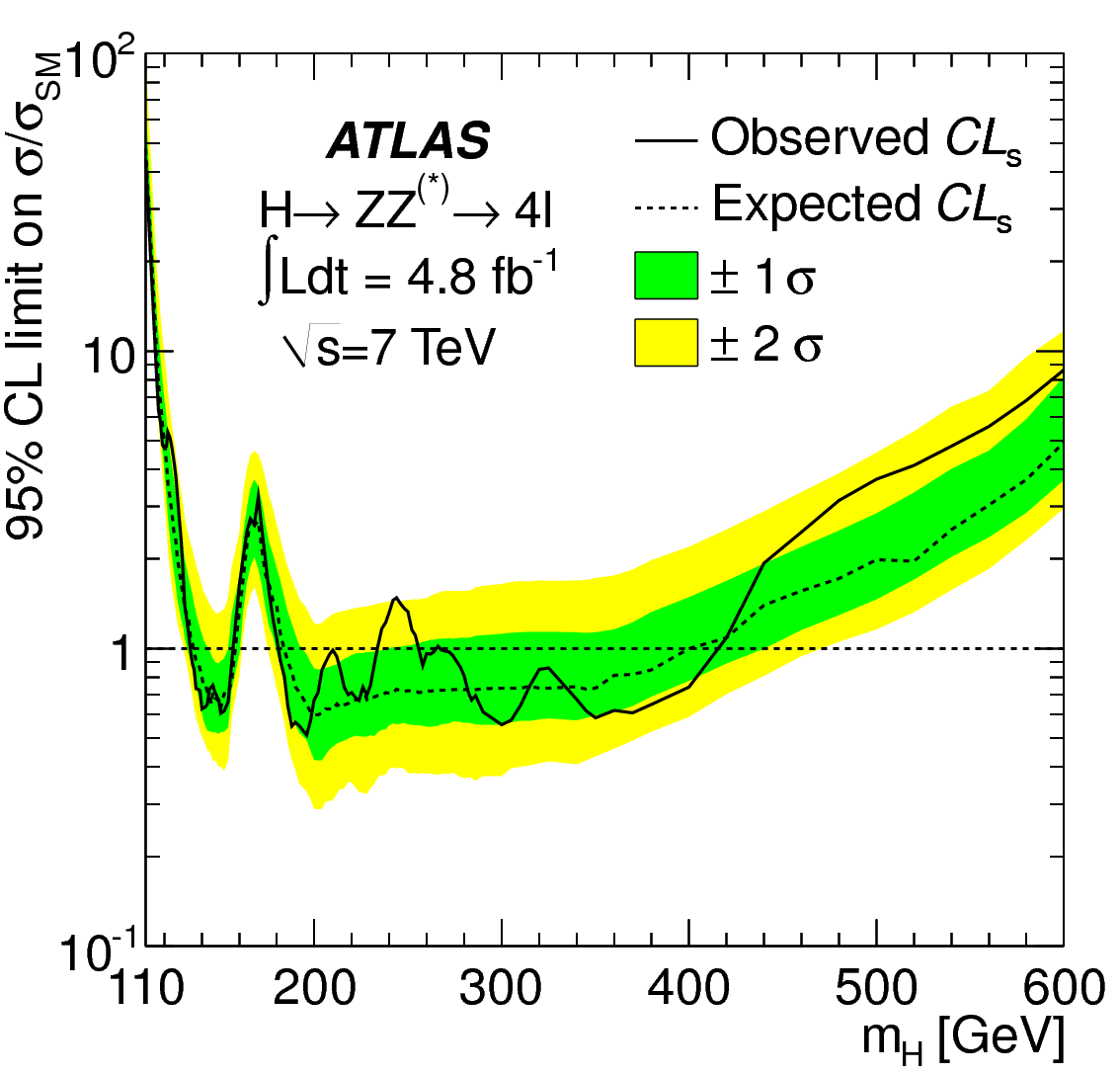,height=65mm}
  \end{center}
  \caption{Left: Invariant four-lepton mass distribution of the selected candidates compared to the background expectation. Right: Observed and expected 95\% CL limits on the SM Higgs boson production in the $H~\to~ZZ^{(*)}~\to~4\ell$ decay channel normalized to the predicted SM cross section as a function of $m_H$.\label{fig:4l}}
\end{figure}

The Higgs boson decays into four-leptons provide a clean signature with a very low level of background. After the lepton isolation and impact parameter cuts, the only remaining background for $m_H\gtrsim$200~GeV is the irreducible $ZZ$ process. For $m_H\lesssim$200~GeV, there is a small additional contribution of the reducible $Zb\bar{b}$, $Z$+jet and $t\bar{t}$ backgrounds estimated from the control data samples. The channel is characterized by high experimental invariant mass resolution of 1.5\% to 2\% up to $m_H$=350~GeV, after which the natural Higgs boson decay width starts to dominate. A high signal efficiency is ensured by a high lepton reconstruction and identification efficiency, as well as the low transverse momentum cut of 7~GeV on the two subleading leptons. The mass of one dilepton pair is required to be compatible with the $Z$ boson mass, while the $m_H$-dependent invariant mass cut is applied on the second lepton pair. The analysis is divided into four sub-channels based on lepton flavours: $4\mu$, $2e2\mu$, $2\mu 2e$ and $4e$, where the first two leptons are assigned to an on-shell $Z$ boson. Figure~\ref{fig:4l} (left) shows the inclusive four-lepton invariant mass distribution after all selection cuts. 
A total of 62$\pm$9 candidates are expected from background processes which is in good agreement with 71 observed candidate events. The corresponding 95\% CL exclusion limit is shown in Figure~\ref{fig:4l} (right). The observed exclusion covers the mass regions from 134~GeV to 156~GeV, 182~GeV to 233~GeV,  256~GeV to 265~GeV and 268~GeV to 415~GeV. The expected exclusion ranges are 136~GeV to 157~GeV and 184~GeV to 400~GeV. The most significant upward deviations from the background-only hypothesis are observed for $m_H$ = 125~GeV with a local significance of 2.1$\sigma$, $m_H$ = 244 GeV with a local significance of 2.2$\sigma$ and $m_H$ = 500~GeV with a local significance of 2.1$\sigma$. Once the look-elsewhere-effect is considered, none of the observed local excesses is significant.

\section{Combined results}

\begin{figure}[!htbp]
  \begin{center}
    \psfig{figure=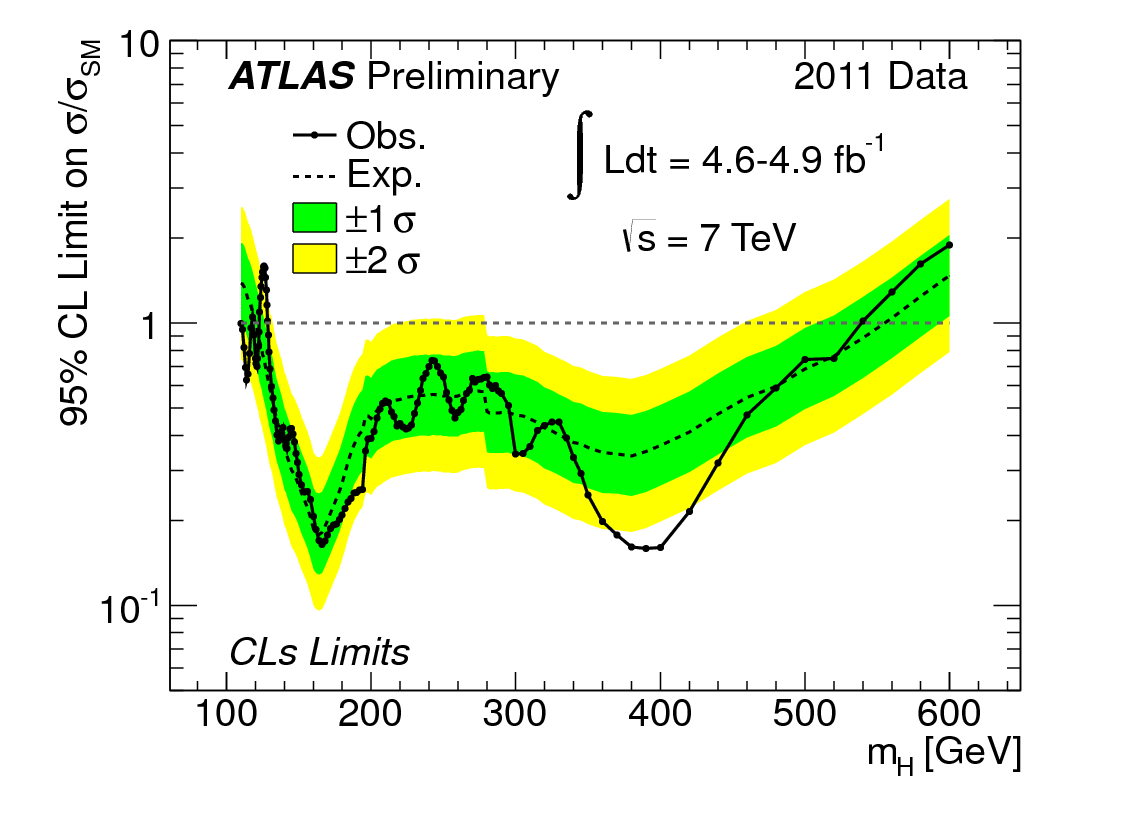,height=60mm}\psfig{figure=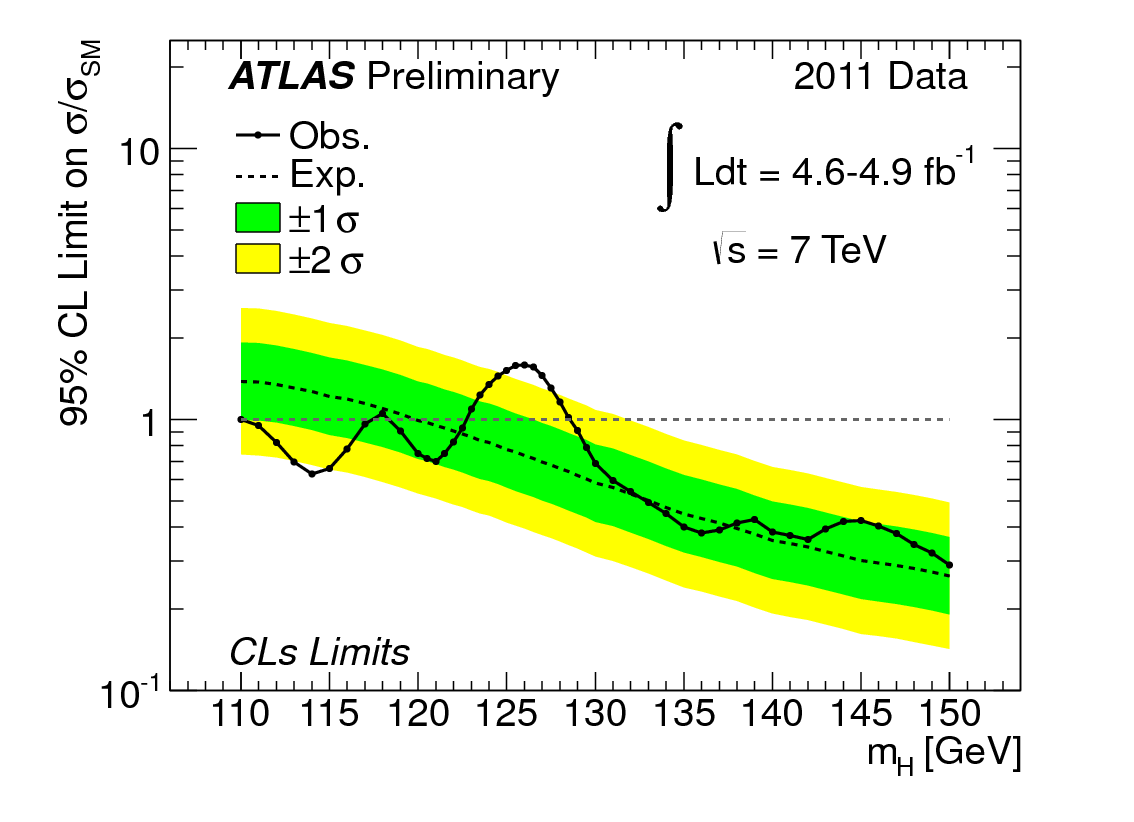,height=60mm}
  \end{center}
  \caption{Combined observed and expected 95\% CL limits on the SM Higgs boson production normalized to the predicted SM cross section as a function of $m_H$ in the entire explored mass range (left) and zoomed in to the low-mass region (right).  \label{fig:comb}}
\end{figure}

The results of the statistical combination of all individual search channels are summarized in Ref.~\cite{combination}. The expected and observed limits obtained are shown in Figure~\ref{fig:comb}. 
SM Higgs boson masses between 120~GeV and 555~GeV are expected to be excluded at the 95\% CL or higher. The observed 95\% CL exclusion regions range from 110.0~GeV to 117.5~GeV, 118.5~GeV to 122.5~GeV, and 129~GeV to 539~GeV. The mass regions between 130~GeV and 486~GeV are excluded at the 99\% CL. 

\begin{figure}[!htbp]
  \begin{center}
    \psfig{figure=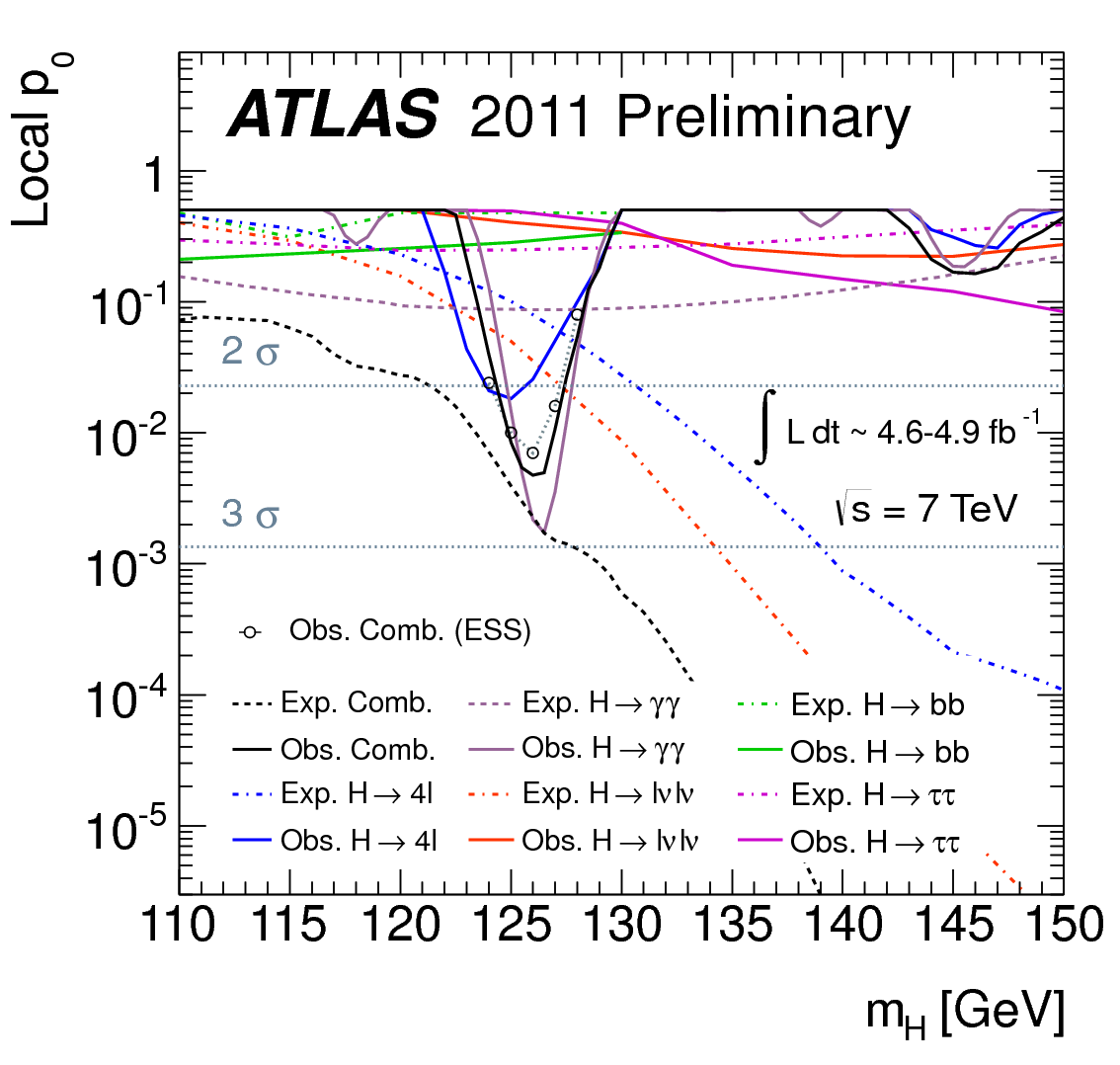,height=60mm}\psfig{figure=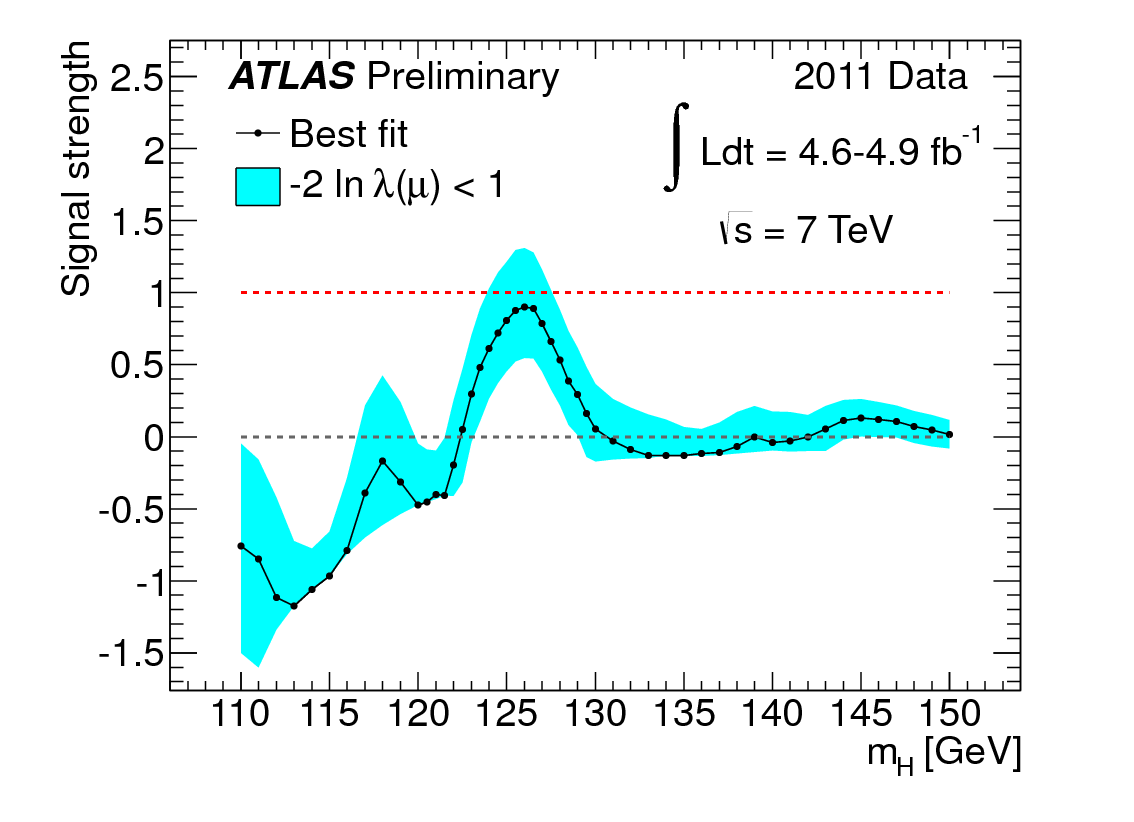,height=60mm}
  \end{center}
  \caption{Left: The local $p_0$-value for individual channels and the combination. The full curves give the observed individual and combined $p_0$-value. The dashed curves show the median expected value under the hypothesis of a SM Higgs boson signal at that mass. Right: The best-fit signal strength $\mu$ = $\sigma$/$\sigma_{SM}$ as a function of the Higgs boson mass hypothesis in the low mass range. \label{fig:comb2}}
\end{figure}

The mass region around $m_H$=126~GeV cannot be excluded due to an observed excess of events compared to the expected background contribution. The $p_0$-values in the given mass region are shown in Figure~\ref{fig:comb2} for individual channels and their combination.  The most significant excesses are observed in the two channels with high mass resolution, $H\to \gamma\gamma$ and $H\to ZZ^{(*)}\to 4\ell$. The largest combined local significance of the observed excess is 2.5$\sigma$, while the expected significance in the presence of a SM Higgs boson with $m_H$=126 GeV is 2.9$\sigma$. The global probability for such an excess to occur anywhere in the entire explored Higgs boson mass region from 110~GeV to 600~GeV is approximately 30\%. The excess corresponds to the best-fit signal strength $\mu$ of
approximately 0.9$^{+0.4}_{-0.3}$, which is compatible with the signal strength expected from a SM Higgs boson at that mass, as shown in Figure~\ref{fig:comb2} (right).

\section{Summary}

The full dataset of $pp$ collision data recorded with the ATLAS detector at the LHC in 2011 has been studied in search for the SM Higgs boson, combining several Higgs boson decay channels in the $m_H$ range from 110~GeV to 600~GeV. A SM Higgs boson is excluded at the 95\% CL in a wide mass range. The exclusion was not possible for $m_H$ around 126~GeV, due to an observed excess of events at the local significance level of 2.5$\sigma$. More data collected in 2012 will be needed to better understand the origin of the described excess.


\end{document}